\documentclass[fleqn,twoside]{article}%
\usepackage{amsmath}
\usepackage{amsfonts}
\usepackage{amssymb}
\usepackage{graphicx}%
\def\be{\begin{equation}}
\def\ee{\end{equation}}
\def\bi{\bibitem}

\begin{document}
\title{Smooth crossing of $w_{\Lambda} = -1$ line in a single scalar field model}
\author{Abhik Kumar Sanyal}
\maketitle
\begin{center}
Dept. of Physics, Jangipur College, Murshidabad, \noindent
West Bengal, India - 742213. \\

\end{center}
\noindent
\begin{abstract}
Smooth double crossing of the phantom divide line $w_{\Lambda}=-1$ has been found possible with a single
minimally coupled scalar field for the most simple form of generalized k-essence cosmological model, in the
presence of background cold dark matter. Such crossing is a sufficiently late time transient phenomenon and does
not have any pathological behaviour.
\end{abstract}
\noindent
\section{Introduction}
Recent analysis of the three year WMAP data \cite{w}, \cite{m}, \cite{p} provides no indication of any
significant deviations from gaussianity and adiabaticity of the CMBR power spectrum and therefore suggests that
the universe is spatially flat to within the limits of observational accuracy. Further, the combined analysis of
the three-year WMAP data with the supernova Legacy survey (SNLS), in \cite{w}, constrains the equation of state
$w_{de}$, corresponding to almost ${74\%}$ of dark energy present in the currently accelerating Universe, to be
very close to that of the cosmological constant value. Moreover, observations appear to favour a dark energy
equation of state, $w_{de}<-1$ \cite{re}. The marginalized best fit values of the equation of state parameter
are given by $-1.14 \le w_{de} \le -0.93$ at $68\%$ confidence level. In case, one considers a flat universe
a-priori, then the combined data leads to $-1.06 = w_{de}= -0.90$. Thus, it is realized that a viable
cosmological model should admit a dynamical equation of state that might have crossed the value $w_{\Lambda}=
-1$, in
the recent epoch of cosmological evolution.\\
So far, it has been administered by Vikman \cite{v} and accepted almost by all \cite{sf}, except perhaps by
Andrianov et-al \cite{a}, and more recently by Cannata and Kamenshchik \cite{b} that smooth crossing of
$w_{\Lambda}=-1$ line is not possible in minimally coupled theories, even through a generalized k-essence
Lagrangian \cite{ke} in the form $\textit{L}=\frac{1}{2}g(\phi)\dot\phi^2-V(\phi)$. It is not difficult to
understand that the standard minimally coupled theory can not go smoothly over to the phantom \cite{cl} domain
without violating the stability both at the classical \cite{rip} and the quantum mechanical levels \cite{qp}
(although it has recently been inferred \cite{on} that quantum Effects which induce the $w_{de} < -1$ phase, are
stable in the $\phi^4$ model). However, Vikman \cite{v}, in particular, argued that transitions from $w_{de}\ge
-1$ to $w_{de}< -1$ (or vice versa) of the dark energy described by a general scalar-field Lagrangian
$(\rho(\phi), \nabla(\phi))$, are either unstable with respect to the cosmological perturbations or realized on
the trajectories of measure zero, even in the presence of k-essence Lagrangian. As a consequence, it has given
birth to further complicated models to establish a smooth crossing. Particularly, it requires hybrid model
composed of at least two scalar fields \cite{l}, one-the quintessence and the other a phantom and is usually
dubbed as quintom model \cite{f}. Others, even complicated models like hessence \cite{h}, holographic dark
energy models \cite{hd}, nonminimal scalar tensor theories of gravity \cite{st}, Gauss-Bonnet gravity \cite{gb},
models with higher order curvature invariant terms \cite{mz} have also been invoked for the purpose. There is
yet another mechanism \cite{dd}, where the crossing is achieved through a dark matter-dark energy interaction in
view of exchange of energy between CDM and quintessence field. As a result the Universe appears to be dominated
by CDM and a crossing dark energy. However,
the same phenomena may be engineered even through the gravitational interaction between dark matter and dark energy.\\
In the present work we have been able to show that the so called phantom divide line corresponding to the state
parameter, $w_{\Lambda}=-1$, can indeed be crossed in a single minimally coupled scalar field model, only by
invoking the most simple form of a generalized k-essence Lagrangian \cite{ke} in the background of baryonic and
non-baryonic cold dark matter. In the model under consideration the state parameter of the dark energy $w_{de}
> -1$, in the absence of the background matter, implying that the field is quintessencial in origin.
It is only through the gravitational interaction the background matter density pushes the dark energy density
down to a
minima and then pulls it up to a local maxima along which the phantom divide line is crossed transiently. \\
The essential feature of the model is a solution of the scale factor in the form,
$a=a_{0}\exp{(\frac{t^f}{n})}$, with $0<f<1$ and $n>0$. Such a solution was dubbed as intermediate inflation in
the nineties \cite{br}. Recently, it has been observed \cite{as} that Gauss-Bonnet interaction in four
dimensions with dynamic dilatonic scalar coupling admits such solution leading to late time cosmic acceleration
rather than inflation at the very early Universe. Under this consequence, a comprehensive analysis has been
carried out \cite{ak} with such solution in the context of a generalized k-essence model. It has been observed
that it admits scaling solution with a natural exit from it at a later epoch of cosmic evolution, leading to
late time acceleration with asymptotic de-Sitter expansion. The corresponding scalar field has also been found
to behave as a tracker field \cite{tr}. Unfortunately, we have not analyzed the behaviour of the state parameter
in the intermediate region. In this work, we show that such solution in the presence of background matter leads
to late time cosmic acceleration with a transient double crossing of the phantom divide line and the Universe is
dominated by the background matter till the second crossing. In section 3, the stability of the model under
linear perturbation has been shown. Thus the model does not develop any pathological features like big-rip or
instabilities at the classical and quantum mechanical level, during the cosmological evolution.\\

\section{The Model}

As mentioned in the introduction, we start with generalized k-essence \cite{ke} Lagrangian in the form,
 \be L = g(\phi)F(X) - V(\phi),\ee
where, $X = \frac{1}{2} \partial_{\mu}\phi \partial^{\mu}\phi$, which, when coupled to gravity may be expressed
in the following most simplest form

\be S=\int d^4x\sqrt{-g}[\frac{R}{2\kappa^2}-\frac{g(\phi)}{2}\phi,_{\mu}\phi^{,~\mu}-V(\phi)+L_{m}], \ee where,
$L_{m}$ is the matter Lagrangian. This is the simplest form of an action in which both canonical and
non-canonical forms of kinetic energies can be treated and a possible crossing of the phantom divide line may be
expatiated. For the spatially flat Robertson-Walker space-time

\[ds^2=-dt^2+a^2(t)[dr^2+r^2\{ d\theta^2+ sin^2(\theta)d\phi^2\}],\]
the field equations in the units $\kappa^2 = 8\pi G = c = 1$ can be expressed as,

\be \dot{H}=-[\frac{1}{2}g\dot\phi^2+\frac{\rho_{m}+p_{m}}{2}],\ee

\be \dot{H}+3H^2=V(\phi)+ \frac{\rho_{m}-p_{m}}{2},\ee

\be g(\ddot\phi+3H\dot\phi)+\frac{1}{2}g'\dot\phi^2+V'=0, \ee  where the dot and the dash represent derivatives
with respect to the time and $\phi$ respectively. In the above equations, $H=\dot a/a$, is the Hubble parameter,
while $p_{m}$ and $\rho_{m}$ stand for pressure and the energy density of the background matter. So, altogether,
we have got three independent equations, viz., (3) through (5), corresponding to six variables of the theory,
viz., $H, \phi, g(\phi), V(\phi), \rho_{m}$ and $p_{m}$. Therefore, we need three physically reasonable
assumptions to obtain complete set of solutions. Our first assumption is to neglect the amount of radiation in
the present day Universe, and to consider the background matter to be filled with luminous along with baryonic
and non-baryonic cold dark matter with equation of state $w_{m}=\frac{p_{m}}{\rho_{m}}=0$. Further, to find a
solution viable for crossing the phantom divide line, we present our second assumption, in the form of the
following ansatz for the Hubble parameter,

\be H=\frac{f}{nt^{1-f}}.\ee with $n>0$ and $f>0$. It is clear that $f=1$, leads to exponential expansion.
However, we choose, $f$ in between, ie., $0<f<1$. Thus the complete set of solutions are

\[a=a_{0}\exp{(\frac{t^f}{n})};\;\;\
\frac{1}{2}g\dot\phi^2=\frac{ f(1-f)}{n t^{(2-f)}}-\frac{\rho_{m}^{0}}{[2a_{0}^3\exp{(\frac{3}{n}t^f)}]};\;\;\
V=\frac{3 f^2}{n^2 t^{2(1-f)}}-\frac{ f(1-f)}{n
t^{(2-f)}}-\frac{\rho_{m}^{0}}{[2a_{0}^3\exp{(\frac{3}{n}t^f)}]};\;\; \] \be\rho_{\phi}=\frac{3 f^2}{ n^2
t^{2(1-f)}} - \frac{\rho_{m}^{0}}{[a_{0}^3\exp{(\frac{3}{n}t^f)}]};\;\;p_{\phi}=\frac{2 f(1-f)}{n
t^{(2-f)}}-\frac{3 f^2}{n^2 t^{2(1-f)}};\;\;\rho_{m}=\frac{\rho_{m}^{0}}{[a_{0}^3\exp{(\frac{3}{n}t^f)}]},\ee
where, $a_{0}$ and $\rho_{m}^0$ are integration constants. The third assumption, which we don't require for the
present purpose, expresses the coupling parameter $g(\phi)$ and the potential $V(\phi)$ as functions of $\phi$.
One can, for example, choose $\phi$ arbitrarily to find different forms of $g$ and $V$, which does not affect
crossing the phantom divide. We shall give one particular form of $g(\phi)$ and $V(\phi)$ at the end. The
effective equation of state $w_{\phi}= \frac{p_{\phi}}{\rho_{\phi}}$ corresponding to the scalar field is now
expressed as,

\be w_{\phi}=a_{0}^3\left(\frac{2nf(1-f)-3f^2~t^f}{3a_{0}^3f^2~t^f-
\rho_{m}^{0}n^2~t^{(2-f)}\exp(-\frac{3}{n}t^f)}\right). \ee The above form of the state parameter $w_{\phi}$
appearing in (8), has been found in an earlier work \cite{ak}, where, we just mentioned that it goes over to
$-1$ value asymptotically. Here, our attempt is to analyze it's behaviour in the interim region. For this
purpose let us express the state parameter $w_{\phi}$ as a function of the red-shift parameter. For
simplification, we choose $a_{0}=1$, without loss of generality. As a result, the constant $\rho_{m}^{0}$,
appearing in equation (7) stands for the amount of matter density present in the Universe at $t=0$. The
red-shift parameter $z$ is defined as,

\[ 1+z=\frac{a(t_{o})}{a(t)}=\exp[{\frac{1}{n}(t_{o}^f-t^f)}],\]
where, $a(t_{o})$ is the present value of the scale factor, while $a(t)$ is that value at some arbitrary time
$t$, when the light was emitted from a cosmological source. Thus, $w_{\phi}$ can now be expressed as,

\be w_{\phi}= \left(\frac{2nf(1-f)-3f^2[t_{o}^{f}-n\ln{(1+z)}]}{3f^2[t_{o}^{f}-n\ln{(1+z)}] - \rho_{m}^{0}~
n^2[t_{o}^{f}-n\ln{(1+z)}]^{\frac{2-f}{f}} \exp{(-\frac{3}{n}[t_{o}^{f}-n\ln{(1+z)}])}}\right).\ee For a
graphical representation of the state parameter versus the red-shift parameter, we need to select a few
parameters of the theory. Firstly, let us choose $f=0.5$ to find $n$. The motivation of choosing the value of
$f$ in the middle is simply to set a comfortable dimension of time for $n^2$ and to obtain a reasonably better
form of the potential $V(\phi)$. Taking the present value of the Hubble parameter $H_{o}^{-1} =
\frac{9.78}{h}~Gyr$, the age of the Universe $t_{o} = 13~Gyr$ and with, $h=0.65$, $n$ can be found from the
ansatz (6) as $n = 0.5 (\frac{H_{o}^{-1}}{\sqrt{t_{o}}})= 2.08$. To estimate the amount of matter density
$\rho_{m}^{0}$ present at the time $t=0$, we take the present value of the matter density parameter
$\Omega_{mo}= 0.26$, and so in view of solution (7),

\[\Omega_{mo}=  \frac{\rho_{mo}}{\rho_{co}}=
 \rho_{m}^{0}(\frac{H_{o}^{-2}\exp{(-\frac{3}{n}t_{o}^f})}{3})=0.26,\] where, $\rho_{mo}$ and $\rho_{co}$ are
the present values of the matter density and the critical density respectively. Thus, we find, $n^2
\rho_{m}^{0}= 2.72.$ With these numbers we have plotted the state parameter versus red-shift parameter in figure
1, which clearly exhibits a smooth double crossing - one from above at $z\approx1.8, ~t\approx2.2$ Gyr and other
from below $z\approx0.44, ~t\approx 8.2$ Gyr (note that in the present model we have started from the value of
the scale factor $a=1$, at $t=0$, corresponding to which the red-shift parameter is approximately $z=4.66$)
which is slightly different from that predicted in view of $\Lambda$CDM model \cite{al}. In figure 2, the matter
density $\rho_{m}$ (thin line) and the dark energy density $\rho_{\phi}$ (thick line) have been plotted against
time in Gyr. It demonstrates that initially the Universe was matter dominated. The crossing from above is
experienced at the minima and that from below at the local maxima of $\rho_{\phi}$. At the local
maxima, $\rho_{\phi}$ overtakes $\rho_{m}$ and the Universe is dominated by dark-energy.\\

{
\begin{figure}
[ptb]
\begin{center}
\includegraphics[
height=2.034in, width=3.3797in] {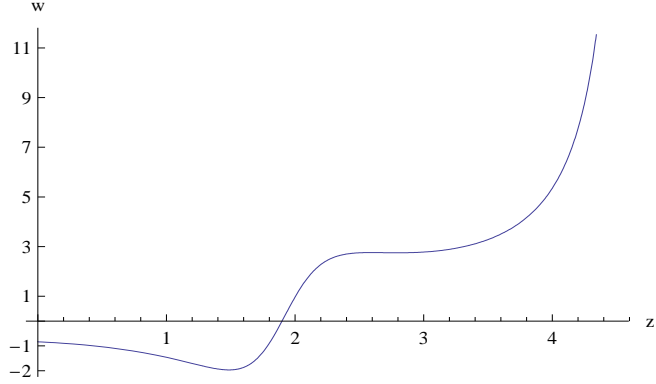} \caption{State parameters $w_{\phi}(z)$ has been plotted against the
red-shift parameter $z$, (with, $f=0.5, h=0.65, t_{0}=13~Gyr.$). Smooth double crossing of the Cosmological
constant barrier is observed at sufficiently later epoch, $z\approx 1.8$ from above and $z\approx 0.44$ from
below.}
\end{center}
\end{figure}

\begin{figure}
[ptb]
\begin{center}
\includegraphics[
height=2.034in, width=3.3797in ] {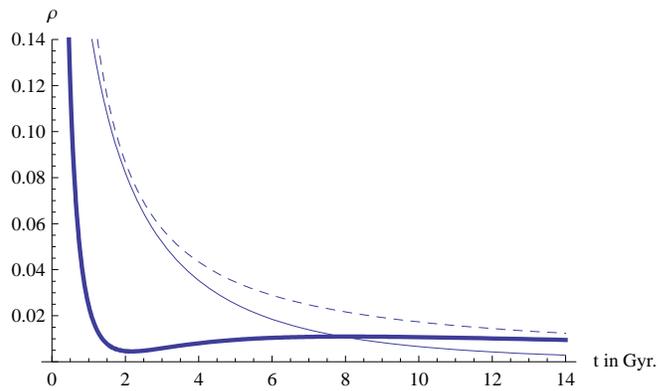} \caption{The tracking behaviour of the scalar field is demonstrated.
Thick and the thin lines correspond to $\rho_{\phi}$ and $\rho_{m}$ respectively. The crossing of the phantom
divide line occurs at the minima and local maxima (where $\rho_{\phi}$ overtakes $\rho_{m}$) of the dark energy
density. Dashed line corresponds to $\rho_{\phi}$ in the absence of background matter. The dark energy is found
to remain subdominant till the second crossing.}
\end{center}
\end{figure}
\noindent
Now we check how far our model fits with the standard $\Lambda CDM$ model. In connection with
$\Lambda$CDM model, the luminosity-redshift relation is,

\be H_{o}dL=(1+z)\int_{0}^{z}{\frac {dz}{\sqrt{0.74+0.26(1+z)^3}}},\ee while in the present model it is,

\be H_{o}dL=\frac{(1+z)}{\sqrt t_{0}}\int_{0}^{z}{[\sqrt t_{0}-n \ln({1+z})]dz}.\ee Further, the expression for
distance modulus is given by,

\be m-M=5log_{10}(\frac{dL}{Mpc}) + 25 = 5log_{10}(DL) + 43\ee where, $m$ and $M$ are the apparent and absolute
bolometric magnitudes respectively and $DL = H_{0}dL$. In view of equations (10) through (12), the distance
modulus - redshift graph has been plotted in Figure 3, which shows almost a perfect fit between the $\Lambda$CDM
and the present models.

\begin{figure}
[ptb]
\begin{center}
\includegraphics[
height=2.034in, width=3.3797in ] {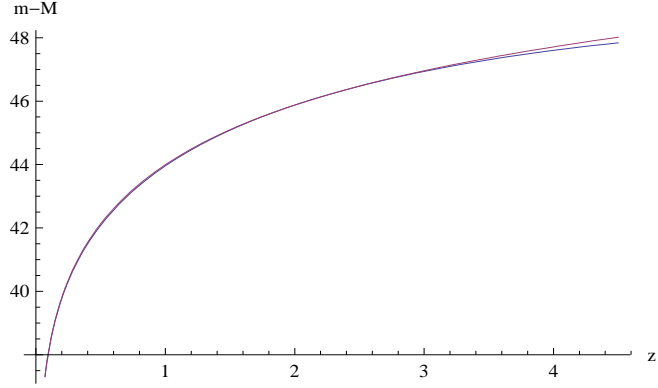} \caption{The fit is almost perfect and the models are nearly
indistinguishable. Only after $z=3.5$, the present model deviates slightly by moving up a little bit from the
$\Lambda CDM$ model.}
\end{center}
\end{figure}
\noindent
Surprisingly, as already mentioned, the crossing depicted here does not depend on a particular form of
the potential $V(\phi)$ or the coupling parameter $g(\phi)$. One can choose $\phi$ arbitrarily to find different
forms of potential and the coupling parameter without affecting the results. For example, under a very trivial
choice $\phi = t$, we have, in view of equation (7),

\be V=\frac{3 f^2}{ n^2\phi^{2(1-f)}}-\frac{ f(1-f)}{ n\phi^{(2-f)}}-
\frac{{\rho_{m}}^{(0)}}{2[a_{0}^3e^{\frac{3}{n}\phi^f}]}, \ee which has already been demonstrated to be a
tracker potential \cite{ak}, and

\be g(\phi)=\frac{ f(1-f)}{ n \phi^{(2-f)}}-\frac{\rho_{m}^{0}}{2a_{0}^3e^{(\frac{3}{n})\phi^f}}.\ee With the
above numerics we also present the plots of $g(\phi)$ and $V(\phi)$, in figures (4) and (5) respectively.

\begin{figure}
[ptb]
\begin{center}
\includegraphics[
height=2.034in, width=3.3797in ] {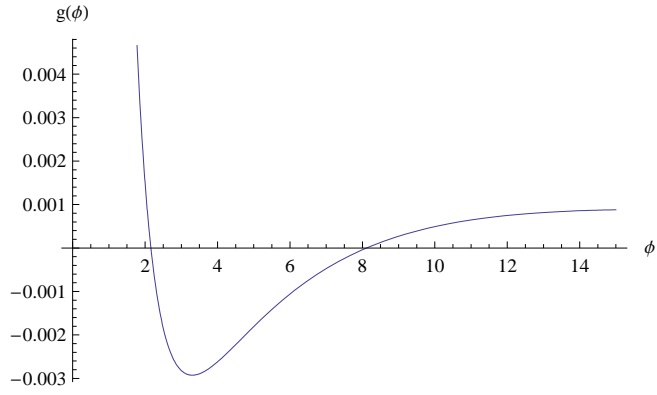} \caption{The figure depicts how $g(\phi)$ smoothly phantomizes and
dephantomizes the model.}
\end{center}
\end{figure}

\begin{figure}
[ptb]
\begin{center}
\includegraphics[
height=2.034in, width=3.3797in ] {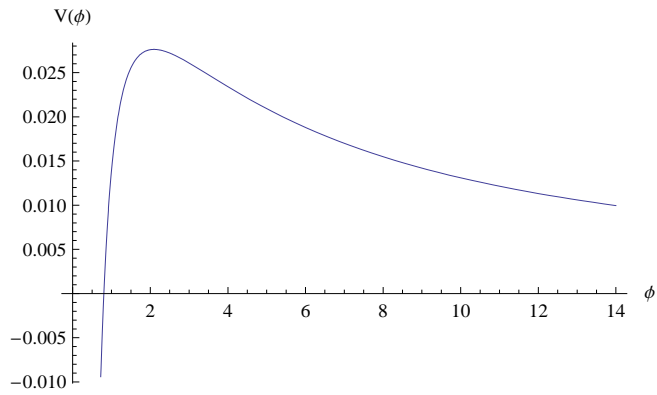} \caption{The potential has a maxima at the first crossing and
thereafter it is almost flat without any peculiarity during the second crossing.}
\end{center}
\end{figure}

\section{Stability Criteria}
Primarily, we note that since there is a dynamical transition of the equation of state from below, so the model
avoids big-rip singularity \cite{rip} and also prevents undesirable quantum mechanical negative energy graviton
and phantom particle production \cite{qp}. As a result classically the model is free from future singularity and
quantum mechanical stability is guaranteed. Further, the velocity of sound \cite{gm}, $\hat{c}_{s}^2 =
\frac{\partial p}{\partial \rho} = \frac{p_{\phi,X}}{\rho_{\phi,X}}$ is always $1$, for the model under
consideration. Now the question is, if the model is stable under appropriate perturbation.\\
First, we recall that in the present model, we have two fluids, one is barotropic $(w_{m} = 0)$ and the other,-a
non-adiabatic scalar field. In view of the solutions (7) and (8) it is clear that in the absence of the
background matter $\rho_{m}^0 = 0$, $g\dot\phi^2 > 0$ and $w_{\phi} = -1 + \frac{2n(1-f)}{3f t^f} > -1$. Thus
the scalar field is of quintessence origin and is not viable of crossing the phantom divide line of it's own.
Though there is no non-gravitational interaction between the two fluids, however they interact strongly under
gravitational influence. In figure 2, the dashed line exhibits the scalar field energy density in the absence of
background. It is the background matter that not only reduces the energy density of the scalar field, but also
generates a minima through which phantomization and a local maxima through which dephantomization take place.
So, one has to consider the system as multi-component Universe \cite{wh} with two fluids where both the
components yield stable fluctuations independently, as in the quintom model. As a result, the composite fluid
with effective pressure $p_{eff} = p_{\phi} + p_{m} = p_{\phi}$, and effective energy density, $\rho_{eff}
= \rho_{\phi} + \rho_{m}$, is also stable.\\
We now first see how the problem encountered with energy-density perturbation raised by Caldwell and Doran
\cite{rrc} is cured. The equation for density perturbation is,

\be \delta \rho_{eff} = \delta (\rho_{\phi}+\rho_{m}) = (-g' F +V')\delta\phi - gF_{,X}\delta X + \delta
\rho_{m} = -(\frac{1}{2}g'\dot\phi^2 + V')\delta\phi - g\delta X + \delta \rho_{m}.\ee In view of the expression
for the state parameter in (8), it is clear that $g\dot\phi^2$ in equation equations (7) vanishes at the
crossing. If we restrict to $\dot\phi^2 > 0$, as in the example cited at the end of last section with $\phi =
t$, then $g(\phi)$ vanishes at the crossing, which has been demonstrated in figure (4). So the coefficient of
$\delta X$ vanishes and the coefficient of $\delta\phi$ also vanishes due to the scalar field equation (5), at
the crossing. However, $\delta \rho_{eff}$ still remains finite at the crossing due to the presence of finite
matter density perturbation
$\delta\rho_{m}$.\\
Next we look for the pressure perturbation equation \cite{pp},

\be\delta p=\hat{c}_{s}^2\delta \rho+3\mathcal{H}(1+w)(\hat{c}_{s}^2-c_{a}^2)\rho\frac{\theta}{k^2},\ee where,
the symbols have their usual meaning. In the expression (16) for pressure perturbation, the adiabatic sound
velocity $c_{a}^2$ diverges at the crossing for one component non-adiabatic fluid and so pressure perturbation
also diverges. In the present model, the expression for the adiabatic sound velocity is

\be c_{a}^2 = \frac{\dot p_{eff}}{\dot\rho_{eff}} = \frac{\dot p_{\phi}}{\dot\rho_{\phi} + \dot\rho_{m}} =
\frac{[3\mathcal{H}w_{\phi}(1+w_{\phi})-\dot w_{\phi}]\rho_{\phi}}{3\mathcal{H}[(1+w_{\phi})\rho_{\phi} +
\rho_{m}]}.\ee Here, dot corresponds to derivative with respect to conformal time and $\mathcal{H}$ is the
Hubble parameter in conformal time. It is observed that at the crossing $c_{a}^2$ remains finite yielding finite
pressure perturbation. So the model is stable under linear perturbations and it appears that everything goes
right with it.

\section{Summary}
First of all it should be admitted that there is nothing crazy in the present model. That the dark energy alone
is not viable of crossing the phantom divide line without classical and quantum mechanical pathological
behaviour has been proved by several authors and that has not been challenged here. One should also admit that a
real cosmic fluid carries baryonic and non-baryonic CDM along with the dark energy and they interact
gravitationally. Vikman \cite{v} studied the behaviour of the k-essence Lagrangian which alone admits crossing
in the background of dark matter, while the dark matter in the present model is a quintessence field with
$w_{\phi} > -1$ in the absence of the background. The fact that quintessence field is viable of crossing the
barrier in the presence of background matter has already been accounted for \cite{dd}, where there exists
non-gravitational interaction between the two. The most interesting feature of the present model is the
observation that though there is no non-gravitational interaction between the background matter and the dark
energy, yet the background matter pushes the energy density of the dark energy (scalar field) to a minima and a
local maxima, around which the transient crossing of the phantom divide line has been experienced. The other
interesting feature is that the phenomena of such transient crossing is independent of any particular form of
the potential. Last but not the least important consequence of the present model, as revealed in figure 2, is
that the matter density had dominant contribution until the second crossing.\\

\noindent
\textbf{Acknowledgement}:{Acknowledgement is due to Prof. Claudio Rubano for some illuminating
discussion.}

\end{document}